\documentclass[prd,letterpaper,twocolumn,floatfix,showkeys]{revtex4}
\usepackage{txfonts}
\usepackage{graphicx}
\newcommand{\gsim}{\raisebox{-0.0em}{$\stackrel{>}{_\sim}$}}
\newcommand{\lsim}{\raisebox{-0.0em}{$\stackrel{<}{_\sim}$}}
\begin{document}
\title{A Natural Explanation for Magnetars}
\author{Dipankar Bhattacharya} 
\affiliation{Raman Research Institute, Bangalore 560080, India}
\author{Vikram Soni}
\affiliation{National Physical Laboratory, K.S. Krishnan Marg, New Delhi 110012, India}
\begin{abstract}
Neutron stars possess some of the strongest magnetic fields known in
the universe. The surface magnetic fields of radio pulsars are 
estimated to be in the range $10^{8}$ to $10^{13}$~Gauss, with 
$10^{12}$~Gauss being the typical value.  Magnetars, a class of 
neutron stars with even stronger magnetic fields, $\sim 10^{15}$~Gauss,
are believed to be ``magnetically powered'' stars, deriving most 
of their radiative luminosity at the cost of their magnetic fields.
The origin of the 
strong magnetic fields of neutron stars, in particular those of magnetars, 
has essentially been an open question for decades. 
In this paper we explore the possibility that a magnetar may owe its
strong field to a magnetized core which, as indicated by certain
equations of state, may form due to phase transitions at high density
mediated by strong interaction within a sufficiently massive neutron
star.  We argue that the field derived from such a core could explain
several inferred evolutionary behaviors of magnetars.
\end{abstract}
\keywords{neutron star; magnetic field; strange matter; pion condensate}
\maketitle

\section{Introduction}
\label{sec:intro}
Magnetars are neutron stars with surface magnetic fields a thousand 
times larger than that of an average pulsar.  However, this is not 
their most distinctive property.  While in pulsars, the loss of 
rotational energy can account for all of the observed radiation, in
magnetars this is not so.
The magnetars exhibit spin down ages $\sim 10^4$ years.
Over this period, they emit a quiescent radiative luminosity 
$\sim 10^{35}-10^{36}$~erg/s. In addition, some of them 
emit repeated flares or bursts of energy typically 
$\sim 10^{42}$--$10^{44}$~erg, and at times much higher: $\sim 10^{46}$~erg
(e.g.\ the very energetic flare of SGR1806-20 on 27 December 2004
\cite{palmer+05,hurley+05}). 
The energy emitted in both quiescent emission and flares far exceeds
the loss in their rotational energy over the same period.  
The only known source of energy for these emissions is their magnetic 
energy \cite{duncan+thompson92,thompson+duncan95}, yet their spin
history appears to present no clear evidence of a decrease in their 
surface magnetic fields with time \cite{thompson+02}.

The range of spin periods in which magnetars have so far been found is
surprisingly narrow, between 5 and 12 seconds.  While the absence of
shorter periods can be attributed to rather fast spin down due to their
strong magnetic field, the upper cutoff in the spin period distribution
implies a sudden ceasing of magnetar activity while their magnetic fields
are still strong.

The distinct radiative property of magnetars indicates that they
have an internal state different from that of ordinary radio
pulsars. During their active period magnetars must have a transient 
store of free energy which dissipates as the observed radiation. 
In this paper we show that if in the collapse of the progenitor to 
a neutron star, the strong interaction creates a strongly magnetic core, 
most of the magnetar behavior can be explained. 
The magnetic field of
the core will initially be screened by the conducting exterior.  As this
configuration relaxes, magnetic free energy is released and drives
the magnetar activity. 
In section~\ref{sec:Bevol} below, we discuss the evolution of the
magnetic field expected in the presence of a strongly magnetised
core.
In section~\ref{sec:eos} we discuss how, in certain equations of 
state of neutron star matter, such a strongly magnetized core could 
naturally arise. Section~\ref{sec:conclusion} summarises our 
conclusions.

\section{Magnetic Field Evolution in Magnetars}
\label{sec:Bevol}
We start with the assumption that in the collapse of the protostar,
a magnetized core is created in the newly formed neutron star with
typical fields at the core surface of $\sim 10^{15}-10^{16}$~gauss.
Any dynamically generated field due to spin alignment via strong
interaction is expected to be of this magnitude, which roughly equals
the product of the nuclear magneton and the baryon number density 
in the neutron star interior.

It is unlikely that the magnetic moment of the core will manifest
itself in a corresponding magnetic field on the neutron star surface
immediately upon its formation.  The material surrounding the core is a 
pre-existing, highly conducting, neutron-proton-electron (npe) plasma. 
As the core undergoes a phase transition to the magnetic phase, this 
surrounding npe plasma will see the magnetic field rise at the core 
boundary in a time scale set by causality, 
$\gsim r_{\rm core}/c \sim 10^{-5}$~sec, which is much longer than its
inverse plasma frequency, $\sim 10^{-21}$~s.  Here $r_{\rm core}$ is the
radius of the magnetic core.  In reality the rise time of the field at the
core boundary may be much longer: the phase transition may start in
small islands, which will then have to come together and merge to form 
a coherent macroscopic structure.  The islands may in principle have
arbitrary magnetic orientation with respect to each other, but any
pre-existing magnetic field, inherited by the star from its progenitor,
will serve to define the orientation of the magnetic moments of these
small regions, and hence also determine the orientation of the net 
magnetic moment of the core.  It therefore appears that the core 
magnetic field will establish itself over a timescale long enough
for the surrounding npe plasma to set up screening currents, 
as per Lenz's law, to prevent the penetration of the magnetic flux 
through it.  In this state, the magnetic field due to the core 
will be squeezed into a tight volume inside the star, around 
$r_{\rm core}$, while an outside observer will have no knowledge of it.  
This magnetic configuration has a much higher energy than a relaxed, 
dipole distribution of the magnetic field outside the core.  Typically, 
if the magnetic flux is confined within a radius $r$, its magnetic 
energy would be proportional to $r^{-1}$.  For a core radius $\lsim 
50$\% of the stellar radius, the excess magnetic 
energy would be of the same order as the energy in the relaxed dipole 
configuration.  

With time, the magnetic field will tend to relax to the dipole configuration
of lower energy, releasing the excess magnetic energy in the process.  This 
will happen because the screening currents have a finite lifetime--they will 
be eventually killed by dissipative processes such as ohmic dissipation and 
ambipolar diffusion \cite{spitzer78}. Time scales for these processes, 
relevant to npe plasma
in the interior of a neutron star, have been worked out by Goldreich and
Reisenegger \cite{goldreich+reisenegger92}.  Their estimates show that for 
typical temperatures in the
neutron star interior, the ohmic dissipation time scale would be very long:
$\gsim 10^{11}$~y.  Ambipolar diffusion will, however, play a very important
role, with a dissipation time scale 
$\sim 5 \times 10^4/(B^2_{16} T^6_{8.5})$~y, where $B_{16}$ is the local 
magnetic field strength in unit of $10^{16}$~G and $T_{8.5}$ is the 
temperature in units of $10^{8.5}$~K, a typical value in the interior of
a young neutron star.  This would suggest that the strong magnetic field 
will emerge at the neutron star surface in the time scale of a few times 
$10^4$~y.  The excess magnetic energy will be dissipated over this period, 
driving an average radiative luminosity $\sim 10^{35}-10^{36}$~erg/s. 
These estimates are in good agreement with the inferred ages and observed 
average luminosities of magnetars.  For a radius of the core about half 
the stellar radius or less, the total free energy stored in the 
initial magnetic configuration would be sufficient to explain the observed 
level of magnetar activity over a duration of $\gsim 10^4$~yr.  The ongoing
dissipation will also serve to keep the stellar interior hot 
enough for ambipolar diffusion to proceed in a reasonably short time scale,
despite its extreme temperature sensitivity.  If the local temperature
tends to rise significantly above $10^{8.5}$~K, cooling via neutrinos
will be dominant and will maintain the temperature at this level.  This
may well happen in the early stages, when the energy is released deep 
into the stellar interior \cite{kaminker+06}.  In this phase the X-ray 
emission from the surface may be relatively faint. The young 
($\sim 1700$~y), high magnetic field ($\sim 4\times 10^{13}$~G) 
radio pulsar J1119-6127 with faint X-ray emission from an unusually hot 
surface may in fact be an example of a star passing through this 
phase \cite{gonzalez+05}.

As the excess magnetic energy is dissipated, the magnetic field at the 
stellar surface will rise, eventually reaching the full dipole value 
corresponding to the core magnetic moment.  This is contrary to what has
been normally assumed in the literature -- that the magnetic field of a
magnetar should decrease with time as a consequence of dissipation of
magnetic energy.  One way of estimating the magnetic fields of magnetars
is from their spin-down rate, attributing the spindown torque to that
due to vacuum magnetic dipole radiation or equivalent.  The same 
technique is used to estimate the magnetic fields of radio pulsars 
\cite{manchester+taylor77}.
The recorded spin-down history of magnetars, however, shows no evidence 
of a secular decrease in their surface magnetic field.  If at all, 
there is a mild evidence in favor of the magnetic field strength increasing
with time \cite{thompson+02}.  A similar trend of increasing field strength has been noted
in certain glitching radio pulsars too \cite{zhang+lin05}.

Magnetars can be classified into two broad categories: Soft Gamma Repeaters
(SGRs) and Anomalous X-ray Pulsars (AXPs) (see Woods and Thompson 
\cite{woods+thompson06} for a recent review).  
SGRs are objects that show relatively frequent burst and flare activity. 
They also possess, on an average, somewhat stronger magnetic fields.
AXPs represent a quieter magnetar population.  Burst activity in them is
rare, and their average field strength is smaller than that of SGRs.
In the conventional scenario, AXPs would be the later evolutionary
products of SGRs \cite{woods+thompson06}.  In our model, SGRs, with stronger 
fields, would
be older than AXPs.  As these neutron stars are born in supernova explosions,
and happen to be just $\sim 10^4$~y old, they are expected to be found 
associated with the remnants of these supernovae.  Association with
Supernova Remnants (SNRs) has indeed been found in several of these cases.
Interestingly, the AXPs are found to be located close to the centers of the
remnants while SGRs are often located far from the centers of the associated
SNRs \cite{thompson+02}.  Given that all neutron stars receive some kick at 
birth, these
observed offsets with respect to the SNR centers could be interpreted as 
the SGRs being older, and hence having had a longer time to drift away 
\cite{thompson+02}.  SGRs have smaller spindown ages ($\equiv P/2\dot{P}$)
than AXPs, but as noted by Thompson, Lyutikov and Kulkarni \cite{thompson+02} 
this may simply reflect an accelerated growth of the surface 
magnetic field in this phase, which is expected to happen as the 
strong magnetic field breaks through the upper crust 
\cite{duncan+thompson92,thompson+duncan95}.

In the time sequence leading to the emergence of the magnetic field at the 
surface of the star, the leading ambipolar diffusion zone moves out from
the interior of the star to the outer surface.  Younger objects, where
the activity is deeper subsurface, are expected to have a relatively
quiet existence, the dissipated heat being conducted out to the surface
and radiated away, powering a quiescent X-ray luminosity.  As the object
ages and the strong field moves through the outer crust, mechanical
disturbances of the crust could be triggered by the magnetic pressure,
leading to glitches and flares.  This trend would qualitatively fit the
difference in the level of burst activity observed in AXPs and SGRs.
The yield strength of the upper crust would be unable to support 
stresses for magnetic field gradient across the crust of $\gsim 10^{13}$~G,
as the maximum stress that the crust can support is estimated to be 
$\lsim 10^{27}$~dyne~cm$^{-2}$ \cite{ruderman91}.
It is therefore likely that in this phase of evolution the magnetic field
will eventually attain its relaxed configuration via several hundreds
of SGR-type bursts.

We have mentioned above that the upper cutoff of the spin period 
distribution of magnetars at $\sim 12$~s implies a sudden ceasing of 
magnetar activity while their magnetic fields are still strong. 
For this there has so far been no natural explanation (see 
\cite{colpi+00} for a discussion).  
In our model, this abrupt stoppage of magnetar activity is in fact 
expected.  Once all the screening currents dissipate away and the 
magnetic field reaches its fully relaxed configuration, there is no more
free energy available in the magnetic field to power the magnetar activity.
Upon reaching this state, the luminosity of the magnetar will plummet and
it will disappear from view.  
As the time scale for the emergence of
the field is inversely proportional to $B^2$ and the spindown torque
on the star is proportional to $B^2$, it would be natural to expect
that at the end of the magnetar activity the spin periods of these
stars will be roughly similar \cite{reisenegger07}.

In the post-magnetar phase the spin periods of these neutron stars
will continue to lengthen, and they may function as long period 
rotation-powered radio pulsars.  Careful surveys sensitive to pulsars
with tens of seconds period may be able to find them.  It is to be
noted, however, that the active lifetimes of pulsars is inversely
proportional to their magnetic field and hence the population of
such long period, high field pulsars is expected to be much smaller
than the garden variety radio pulsars with teragauss surface field.

\section{Origin of magnetic cores in neutron stars}
\label{sec:eos}
In the discussion above we have assumed the presence of a strongly
magnetized core in the neutron stars that act as magnetars. There are 
several scenarios in which such core magnetic fields can be created
dynamically via a phase transition at high density for strongly 
interacting matter. To name a couple of possibilities,
\begin{enumerate}
\item In classic old fashioned nuclear high density ground states,
      large macroscopic magnetic moment can be generated by the
      alignment of nuclear spins, often accompanied by pion condensation 
      (e.g. \cite{baym77,dautry+nyman79}).
\item It is considered more plausible that at typical neutron star
      central densities ($\sim 5-6$ times nuclear density) we expect
      the nucleons to dissolve into quark matter.  We have found that
      the likely ground state of such matter at these densities is 
      a neutral pion condensate \cite{soni+bhattacharya04,kutschera+90}.
      Such a state carries a large magnetic moment {\em and
      hence can give rise to a strongly magnetized core}.
\end{enumerate}
Both these cases involve a phase transition to a condensate, with 
condensation energies of order several tens of MeV per baryon
(see, e.g., ref.~\cite{sadzikowski+broniowski00}), corresponding to
transition temperatures in excess of $10^{11}$~K.  The interior 
temperature of a neutron star would fall well below this
within a day after formation 
\cite{pethick92,yakovlev+pethick04}, resulting in the formation of a 
magnetic core.

Several other equations of state of quark matter have been considered in the 
literature.  One that has attracted much interest of late is that with a
diquark superconductivity
\cite{alford+98,rapp+98}.  A transition to such a state is expected at high 
density ($\sim 10$ times nuclear density).
Ferrer and de la Incera \cite{ferrer+incera06} have considered a scenario 
where strong 
magnetic fields can be produced by the generation of gluon vortices in the 
colour superconducing cores of neutron stars. This mechanism achieves
an amplification of a pre-existing strong ($\sim 10^{17}$~G) field.
On the other hand, the magnetic pion condensate mentioned above can 
spontaneously generate magnetar strength fields (see fig.~1).

We now present some considerations that will demarcate regions in the
parameter space that can support quark matter Pion Condensed cores and
those which are purely nuclear.

The magnetic 
pion condensed ground state of quark matter from an effective 
 intermediate chiral lagrangian was first considered by Kutschera 
et al \cite{kutschera+90} for two-flavor quark matter. We have generalised
 the effective lagrangian, L, to 3 flavours, which can be used over a large range
 of densities, from nuclear matter
to quark matter. We have also, for the first time, fixed all
the coupling parameters of, L,  from low energy hadronic data (see \cite{soni+bhattacharya06} 
for details).
One of the parameters is the (tree-level) mass of the Sigma particle which is related to
$\pi \pi$ scattering, and is estimated to be around $m_{\sigma} \sim 800$~MeV
\cite{abdel-rehim+03}.
 
We have carried out this calculation for charge neutral, three-flavor
quark matter in $\beta$ equilibrium.  This includes
strange quarks as well as one-gluon exchange interactions. We have shown 
in an earlier work \cite{soni+bhattacharya04} that if $m_{\sigma}$ exceeds 
$\sim 700$~MeV, 
then according to this mean field theory, the quark matter, rather than being 
in the conventional strange quark matter (SQM) state \cite{witten84}, 
would find it energetically favorable to organize itself 
into a chiral $\pi^0$ Pion Condensate (PC) which is spin polarized.
Our calculations \cite{soni+bhattacharya06} show that
in the PC state that is realized, at the densities of interest, in the
presence of the $\pi^0$ condensate the strong interaction causes the u-quark 
quasiparticles to align their spins along the condensate wave vector
$\vec{q}$ and the spins of d-quark quasiparticles to align the opposite way.  
The two species being oppositely charged, their magnetic moments then
add, giving rise to local magnetic fields of order $10^{16}$~G. 
We estimate the magnetic moment density as $e\hbar(2S_u+S_d)/3mc$,
where $S_u$ and $S_d$ are the spin densities of u and d quarks 
respectively. The quark mass $m$ is set to $g<\sigma>$, $g$ being
the Yukawa coupling constant between the meson and the quark fields, which
is set to 5.4 (see ref.~\cite{birse+banerjee84}) by fixing the nucleon
mass to be 938 MeV.  

The equation of state of this PC matter has been derived and presented 
in \cite{soni+bhattacharya06}, where we also discuss the structure of 
neutron stars that would result when the PC quark matter resides in the
core. The outer part of the star is made up of nuclear matter, for which
we adopt the equation of state derived by Akmal et al \cite{akmal+98}
(APR), used commonly in the literature.  The boundary between the nuclear 
and the quark matter PC phase is treated as a first order phase transition. 
The main results are as follows.

\subsection{The quark matter PC core regime (750~MeV$< m_{\sigma}<$850~MeV)}
\begin{enumerate}
\item In this model, the parameter $m_{\sigma}$ is constrained to be above
      750~MeV, to ensure that the two-flavor ground state does not fall
      below the nuclear ground state \cite{soni+bhattacharya04}.
\item The maximum mass of neutron stars with PC cores works out to be
      $M_{\rm max,PC}\sim 1.6\;M_{\odot}$, as in most other cases of neutron 
      stars with quark matter core.  In contrast, the maximum mass of a
      purely nuclear star with APR equation of state is 
      $\sim 2.2\;M_{\odot}$. 
      
      The equation of state for PC quark matter employed by us includes
      gluon interactions as given by Baym \cite{baym77,baym+chin76}.
      Other prescriptions for this exist in the literature, and
      some attempts have led to the mass limit of a star with a quark core 
      being raised to as high as $\sim 2\;M_{\odot}$ \cite{alford+06,alford+05}.
\item The PC quark matter core occurs only if the mass of the star exceeds 
      a threshold
      value $M_{\rm T}$.  Below this threshold mass the star is composed
      entirely of nuclear matter. The threshold mass is a function of the 
      parameter $m_{\sigma}$. With increasing $m_{\sigma}$, the density of 
      phase transition between the APR and PC phases rises and so does 
      $M_{\rm T}$. For an $m_{\sigma}$ of 815~MeV, $M_{\rm T}$ works out to be 
      $1.32\;M_{\odot}$. A value of $m_{\sigma} > 850$~MeV results in 
      $M_{\rm T}$ rising above $M_{\rm max,PC}$ and thereby precludes 
      the formation of a PC core.  Neutron stars with quark matter 
      PC cores can therefore 
      exist only if $m_{\sigma}$ is in the range 750--850~MeV.
\end{enumerate}

To find the contribution of the PC core to the stellar magnetic field
we estimate the magnetic moment $\mu$ by integrating the magnetic moment 
density 
over the volume of the PC core.  The effective polar field strength at the 
stellar surface in fully relaxed configuration may then be computed as:
$B_{\rm s} = 2\mu/R^3$, where $R$ is the stellar radius.  The result, for
two different values of $m_{\sigma}$, are shown in fig.~1.  As can be seen, 
$B_{\rm s}$ has a typical value of $\sim 10^{15}$~G, similar to the fields
encountered in magnetars.  The field strength increases with the mass of 
the star above the threshold mass, and reaches $\sim (2-3)\times 10^{15}$~G 
near the maximum mass.

\begin{figure}
\includegraphics[angle=270,width=\columnwidth]{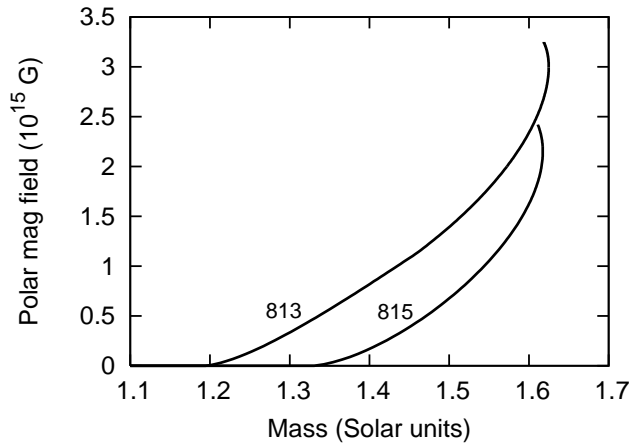}
\caption[x]{The polar magnetic field strength at the surface of a neutron 
star containing pion-condensed quark matter core.  The surface field strength 
is estimated for a fully relaxed dipole configuration corresponding to the 
magnetic moment of the core, shown for two values of the parameter 
$m_{\sigma}$, 813 and 815 MeV. The core appears above a threshold mass 
and if present, could generate surface fields of strength similar to those 
seen in magnetars.}
\end{figure}

If magnetars do indeed owe their magnetic fields to such pion cores, their
masses must exceed $M_{\rm T}$.  Ordinary pulsars, with
much weaker magnetic fields $(\sim 10^{12}~G)$, may then represent a
neutron star population with masses less than $M_{\rm T}$, their magnetic 
fields being either inherited from their progenitors \cite{woltjer64}, 
or amplified in the crust by thermoelectric processes \cite{blandford+83}.  
This can be tested if the masses of magnetars can be measured, perhaps
when magnetars in binary systems are found.  Masses of several pulsars
are known, and in most cases lie close to $\sim 1.4\;M_{\odot}$.

\subsection{The Purely Nuclear Matter star regime ($m_{\sigma}>$850~MeV)}
At high density
(above $\sim 2- 3$ times nuclear density) the nuclear matter itself can 
have a neutral pion condensate \cite{baym77,dautry+nyman79},
in which nuclear spins, aligned due to the
presence of the neutral pion condensate, can contribute a strong 
magnetic field.
For $m_{\sigma} > 850$~MeV the quark matter - nuclear matter transition 
density rises such that
only fully nuclear stars are stable and can continue to
exist. Since it is possible in a purely nuclear matter star at
high enough density to have nuclear PC core, we then have a nuclear matter
analogue for the lower threshold mass,  $M^N_{\rm T}$ for nuclear magnetars
with a new and higher maximum mass $M^N_{max}$ of 
the star (as purely nuclear EOS's are stiffer).

We point to this possibility as
one object, PSR~J0751+1807 has recently been found to be relatively 
heavy, with a mass of $2.1^{+0.4}_{-0.5} \;M_{\odot}$ \cite{nice+05}.  
While this is still marginally consistent with $M_{\rm max,PC}$, if
future measurements confirm it to be larger than $\sim 1.6\;M_{\odot}$
then the existence of pure quark matter PC core in neutron stars would be
ruled out. Core magnetic fields through spin alignment may still be
produced, but perhaps in a pion-condensed nuclear phase 
\cite{dautry+nyman79} or a mixed phase \cite{glendenning92} and the
rest of the magnetar scenario would remain the same as above. 

This discussion was to indicate that magnetars are possible both with
a PC quark matter core or with a PC nuclear core.  However if a neutron
star with mass above $\sim 1.6\;M_{\odot}$ is found then this would 
preclude pure quark matter core for all stars.
However, as we have indicated above, it is possible that a different 
treatment of gluon interactions may raise the mass limit of stars with 
PC quark cores, as has been obtained in some equations of state involving 
quark matter \cite{alford+06,alford+05}.

\subsection{Magnetar by birth or accretion}
Neutron stars with a large mass could result either,(i) from the core
collapse of a rather massive star or (ii) by heavy accretion onto a neutron
star in a binary system.  In either case, if the final mass exceeds
$M_{\rm T}$, a magnetic pion core will form.  Will one expect to see
a magnetar in all such situations?  The answer depends on the details
of the thermal structure in the neutron star interior.  

(i) A newly born
neutron star in a stellar collapse has a very hot interior, facilitating
ambipolar diffusion and allowing the strong field to emerge at the surface
in a short time, as discussed above.  The more massive the neutron star, 
the larger will be the size of the magnetic core and the quicker will the 
strong field emerge to the surface.  It is interesting to note that possible 
magnetar activity has been indirectly inferred in the $\sim 350$~yr old 
supernova remnant Casseiopeia A \cite{krause+05}, which originated in 
the explosion of a star as massive as $\sim 25$~$M_{\odot}$ \cite{vink04}.  
Most importantly, recent observations \cite{muno06} have uncovered several
cases of association between magnetars and massive star clusters indicating
that the progenitors of these neutron stars were more massive than 
$\sim 30\;M_{\odot}$.  A similar conclusion is drawn in the case of 
AXP 1E 1048.1-5937 which is associated with a stellar wind-blown shell 
of neutral hydrogen \cite{muno06}.

(ii) A neutron star accumulating matter
via accretion, on the other hand, is old and its interior is relatively
cold.  Heating due to the accretion process is not expected to raise the
interior temperature above $\sim 10^8$~K \cite{zdunik+92}.  The extreme
temperature sensitivity of the ambipolar diffusion rate will then
delay the emergence of the field at the surface, perhaps for such a
long time that the magnetar property would never be visible.  
This may be the reason why the surface magnetic field PSR~J0751+1807
remains low \cite{nice+05} despite its mass growing to a large value. 

\section{Conclusions}
\label{sec:conclusion}
We have found that in certain allowed equations of state of neutron
stars a strongly magnetized pion condensed core occurs at high density,
for neutron star masses near the maximum mass.
We have argued that such stars would have properties
remarkably similar to the observed magnetars.  The magnetic field of the core
would initially be squeezed deep into the star due to screening currents
set up in the material surrounding the core.  The free energy in this field
configuration would be released over a time scale of $\sim 10^4$~yr and
power the magnetar activity. Along with this, the magnetic field at the
surface will rise, possibly turning Anomalous X-ray Pulsars into Soft
Gamma Repeaters. If this model for magnetars is correct, then
magnetars would be expected to have masses higher than a certain threshold, 
and would therefore be expected
to be associated with supernova explosions from relatively massive stars.
Neutron stars with magnetic cores may also be produced by heavy accretion 
onto normal neutron stars in binary systems, although it seems less likely
that the magnetic field will emerge to the surface quickly enough to 
exhibit magnetar activity in this case.  If a magnetar in a binary system 
is found, several of these predictions could be tested.

\section*{Acknowledgments}
We thank S. Bonazzola, B. Carter, V. Kaspi, S.R. Kulkarni, N. Prasad, 
V. Radhakrishnan, A.R.P. Rau, A. Reisenegger, C.S. Shukre, S. Sridhar, 
K. Subramanian and 
L. Woltjer for discussions and valuable comments.  A compendium of magnetar 
observations, compiled by J. Belapure during summer 2005, has been helpful in 
guiding our thoughts.  VS thanks Raman Research Institute for hospitality 
during the course of this work, and the Indo French center for the 
Promotion of Advanced Research for providing the opportunity to discuss
with Brandon Carter at Observatoire de Meudon and J. Y. Ollitraut, 
Marc Chemtob, R. Schafer and J.P. Blaizot at CEA Saclay.  We thank the
anonymous referees for constructive comments that led to several 
improvements in the paper.

\end{document}